\documentclass[authoryear]{elsarticle}
\usepackage{setspace}   %Allows double spacing with the \doublespacing command
\usepackage{rotating}
\usepackage{float}
\usepackage{hyperref}
\usepackage{amsmath}
\usepackage{color}
\newcommand{\beginsupplement}{%
        \setcounter{figure}{0}
        \renewcommand{\thefigure}{S\arabic{figure}}%
     }

\begin{document}
\begin{frontmatter}
\begin{abstract}
\noindent The human brain is a complex network of interconnected brain regions organized into functional modules with distinct roles in cognition and behavior. An important question concerns the persistence and stability of these modules over the human lifespan. Here we use graph-theoretic analysis to algorithmically uncover the brain's intrinsic modular organization across multiple spatial scales ranging from small communities comprised of only a few brain regions to large communities made up of many regions. We find that at coarse scales modules become progressively more segregated, while at finer scales segregation decreases. Module composition also exhibits scale-specific and age-dependent changes. At coarse scales, the module assignments of regions normally associated with control, default mode, attention, and visual networks are highly flexible. At fine scales the most flexible regions are associated with the default mode network. Finally, we show that, with age, some regions in the default mode network, specifically retrosplenial cortex, maintain a greater proportion of functional connections to their own module, while regions associated with somatomotor and saliency/ventral attention networks distribute their links more evenly across modules.
\end{abstract}

\title{Functional brain modules reconfigure at multiple scales across the human lifespan}
\author{
Richard F. Betzel$^{1,2,}\corref{corr}$,
Bratislav Mi\v{s}i\'{c}$^1$,
Ye He$^4$,
Jeffrey Rumschlag$^1$,
Xi-Nian Zuo$^4$,
Olaf Sporns$^{1,3*}$
}
\address{$^1$ Indiana University, Psychological and Brain Sciences, Bloomington IN, 47405, USA}
\address{$^2$ University of Pennsylvania, Department of Bioengineering, Philadelphia, PA, 19104, USA}
\address{$^3$ Indiana University, Network Science Institute, Bloomington IN, 47405, USA}
\address{$^4$ Key Laboratory of Behavioral Science and Magnetic Resonance Imaging Research Center, Institute of Psychology, Chinese Academy of Sciences, Beijing, China}

\cortext[corr]{corresponding author: \texttt{rbetzel @ seas.upenn.edu}}


\end{frontmatter}

%\linenumbers
\section*{Introduction}
\doublespacing
One of the hallmark properties of complex networks is that they can be analyzed at multiple levels, ranging from that of individual nodes and edges to global descriptions of the whole network. Between these two extremes lie intermediate levels at which networks can be characterized based on decompositions of the network into groups of nodes. At these levels a network can be described in terms of its community structure \citep{porter2009communities, fortunato2010community, newman2012communities}, where a community (also called a ``module'') refers to a densely interconnected set of nodes sparsely connected to the rest of the network \citep{sporns2015modular}. Communities have different meanings depending upon the class of network one considers. In social networks, for example, they represent work groups \citep{leskovec2008statistical} or online virtual communities \citep{traud2011comparing} of individuals or actors, whereas in biological networks communities might correspond to groups of proteins \citep{guimera2005functional} or other cellular components \citep{ravasz2002hierarchical} that perform similar functions.

The flexibility of the network model has made it appealing to many areas within the biological sciences. In neuroscience recent technological advances have made it possible to represent the anatomical and functional interactions among brain regions as complex networks \citep{bullmore2009complex, rubinov2010complex}. The functional connectivity (FC) between two brain regions expresses the statistical dependence of their neurobiological activity, usually operationalized as a correlation \citep{friston2011functional}. The set of all pairwise correlations can be arranged to form a square matrix, which specifies a functional brain network. Like other complex networks, functional brain networks exhibit community structure and can be partitioned into groups of mutually correlated brain regions, which display characteristic topographic patterns at rest \citep{yeo2011organization, power2011functional, doucet2011brain} and that reconfigure in response to task demands \citep{cole2014intrinsic, kitzbichler2011cognitive, stanley2014changes, liang2015topologically} and with learning \citep{bassett2015learning}. These communities, often referred to as ``intrinsic connectivity networks'' (ICNs) have distinct cognitive-behavioral fingerprints \citep{smith2009correspondence, crossley2013cognitive} and have also been implicated in neuro-pathology and disease \citep{alexander2010disrupted, fornito2015connectomics}.

Functional brain networks are simultaneously persistent and flexible across the human lifespan. Even in preterm infants, many of the features that typify adult brain networks are already evident, including proto-ICNs for visual, auditory, and somatosensory systems \citep{fransson2007resting, smyser2010longitudinal}. Through early childhood and adolescence, these systems undergo refinement as short-range connections are gradually replaced by longer connections, so that by early adulthood we find recognizable, distributed functional systems \citep{fair2007development, fair2008maturing, fair2009functional, kelly2009development, supekar2009development, gu2015emergence}. Advanced aging, on the other hand, is generally accompanied by a weakening of functional connections, especially long-distance anterior-posterior connections \citep{ferreira2013resting}. Aging may also disproportionately affect the default mode network, as both local \citep{tomasi2012aging} and long-range connections weaken \citep{andrews2007disruption}.

An important question is how the brain's community structure changes with age. Most studies that try to address this question make comparisons between specific age ranges, for example preterm infants and adults \citep{van2014neonatal} or adults of different age groups \citep{meunier2009age, geerligs2014brain}. More recently, several papers have investigated changes in community structure from childhood to senescence, treating age as a continuous variable \citep{cao2014topological, betzel2014changes, chan2014decreased}. Overall, these studies converge in their findings and suggest that communities become less segregated with age, especially with advanced aging. In most of these studies, communities were defined ahead of time based on canonical representations of ICNs or uncovered using community detection methods that deliver a single partition. This entails two important limitations. First, this approach results in a ``definitive'' description of community structure at a particular scale, where scale refers to the size and number of communities. Such an approach does not address the possibility that communities exist over a range of scales \citep{betzel2013multi} (i.e. multi-scale community structure), or are organized hierarchically \citep{meunier2009hierarchical}. Second, this approach assumes that community structure remains fixed across age groups, implying that functional communities (subsystems) cannot dissolve, change their boundaries, and that no novel communities emerge across the lifespan.

The aim of the present study is to investigate how the brain's community structure evolves over the course of the human lifespan. We construct representative functional networks for different age groups from a cohort of 316 participants covering a large portion of the human lifespan, treating each network as a layer in a multi-layer network representation. Using community detection methods, we algorithmically resolve communities across a range of scales. We show that the modularity of functional brain networks, which measures the degree to which communities are segregated from one another, follows a scale-specific trajectory across the lifespan: at coarse scales, communities become more modular (more segregated), while communities defined at finer resolutions become less modular (less segregated). We also show that community structure is not fixed across the lifespan and that brain regions move from one community to another with the greatest frequency occurring around young adulthood. Finally, we show that brain regions' participation coefficients, which measure the extent to which their links are distributed uniformly across communities, evolve with age. These results suggest that the process of lifespan development is associated with changes in the modular organization of brain functional connectivity at multiple scales.

\section*{Methods}
\subsection*{Data acquisition and processing}
The NKI-Rockland Sample (NKI-RS) is an ongoing project that aims to generate a large ($N>1000$) cross-sectional dataset, where the quasi-experimentally-manipulated variable is the participant's age at the time of data collection \citep{nooner2012nki}. This study was approved by the NKI review board and all participants provided informed consent prior to data collection. As part of the data collection process, each participant completed one anatomical scan, one diffusion structural scan and three resting-state functional MRI (rfMRI) scans that varied in terms of TR time, voxel size, and scan duration: 1) TR = 2,500 ms, voxel size = 3 mm, scan duration = 5 min; 2) TR = 1,400 ms, voxel size = 2 mm, scan duration = 10 min; and 3) TR = 645 ms, voxel size = 3 mm and duration = 10 min. We analyzed the fastest multiband imaging data, which appeared superior to the other acquisitions in terms of reproducibility of rfMRI \citep{zuo2014test}. More details on these data are publicly accessible via the FCP/INDI website
(\url{http://fcon_1000.projects.nitrc.org/indi/enhanced/index.html}). All image data were preprocessed using the Connectome Computation System (CCS) pipeline. The preprocessing strategy included discarding the first several volumes (10 seconds), removing and interpolating spikes that arise from either hardware instability or head motion, slice-time correction, image intensity normalization, and removing the effect of physiological noise by regressing out twenty-four parameters from a motion model \citep{yan2013comprehensive, satterthwaite2013improved} as well as nuisance variables such as white matter and cerebrospinal fluid signals, along with both linear and quadratic trends. Details of the image preprocessing steps are described in \cite{xu2015connectome}. In total, we processed data from 418 individual participants. The quality control procedure in the CCS excluded 64 participants due to their low-quality multimodal imaging datasets, which met at least one of the following criteria: (1) failed visual inspection of anatomical images and surfaces; (2) mean frame-wise displacement $>$ 0.2 mm; (3) maximum translation $>$ 3 mm; (4) maximum rotation $>$ 3$^{\circ}$; or (5) minimum cost of boundary-based registration (a measure of image registration quality) $>$ 0.6. Additionally, thirty-two participants were excluded from subsequent analyses because of clinical diagnoses as defined by DSM-IV or ICD10 or incompleteness of the multimodal imaging datasets. Finally, six participants were excluded, as they also participated in the pilot stage of data collection. This leads to a final lifespan sample of 316 healthy participants. For the age distribution of participants see Figure \ref{fig:figs1}.

\subsection*{Network construction}
For each of the $N=316$ participants,we constructed a weighted and signed functional connectivity matrix, $\mathbf{W}=[W_{ij}]$, whose elements denote the connection weights among pairs of $n=113$ cortical regions of interest, which were represented as nodes in our networks. Regions of interest were defined based on a sub-division of the system assignments of \cite{yeo2011organization} such that each node was anatomically isolated from other regions with the same system assignment and was separated by system boundaries from regions assigned to other systems. The weight of the connection between nodes $i$ and $j$ was given by $W_{ij}=\frac{1}{T - 1}\sum_{t=1}^T z_i (t) \cdot z_j (t)$, where $z_i (t)=\{z_i (1),\ldots,z_i (T)\}$ was the standardized (i.e. zero mean, unit variance) fMRI BOLD time series for region $i$. To study age-related changes in functional brain networks we constructed representative matrices for different age groups. Briefly, this process entailed assigning each participant to one of $K$ non-overlapping age groups. To facilitate statistical comparisons, the boundaries of age groups were chosen so that each group contained approximately the same number of individual participants. For each group, $r \in \{1,\ldots,K\}$, we generated a composite matrix, $\mathbf{W}_r$, by selecting, at random, half the subjects assigned to that group and averaging their connectivity matrices. We then treated each group's composite matrix as a layer in a multi-layer network, $\mathcal{W}=\{\mathbf{W}_1,\ldots,\mathbf{W}_K\}$. We repeated this process 500 times, thereby generating an ensemble of multi-layer networks from many sub-samples of the $N=316$ participants. The analyses described in the main body of the text were carried out over this ensemble of multi-layer networks with $K=5$ age groups or layers. The range of ages included in each group was: 8.3-22.4, 22.4-41.0, 41.0-51.4, 51.4-62.8, and 62.8-83.4 years). In the Supplement we explore the robustness of our results with $K=4,6$, and $7$ age groups.

\subsection*{Single-scale modularity}
The primary focus of this study was on the concept of communities (or modules) in functional brain networks. In practice, real-world networks are usually too big or too complex to identify modules by simple inspection. Finding communities in complex networks requires algorithmic ``community detection'' tools \citep{fortunato2010community}. The range of methods available for detecting communities is broad \citep{palla2005uncovering, rosvall2008maps, ahn2010link, lancichinetti2011finding, zhang2014scalable}, though the most common approach involves dividing a network's nodes into non-overlapping clusters based on the partition that maximizes the ``modularity'' quality function \citep{newman2004finding}:

\begin{equation}
Q=\sum_{ij} B_{ij}\delta(g_i,g_j)
\end{equation}

where $B_{ij}=W_{ij}-P_{ij}$ is the actual weight of the connection between nodes $i$ and $j$ minus the expected weight, $P_{ij}$. The matrix $\mathbf{B}=[B_{ij}]$ is known as ``the modularity matrix''. Thus, modularity maximization aims to assign each node to a cluster, $g_i \in \{1,\ldots,C\}$, so that the positive elements of $B_{ij}$ fall within clusters and $Q$ achieves as large a value as possible. These clusters are then treated as estimates of the network's communities. Clusterings that result in greater modularity scores are generally considered to be of higher quality (Figure \ref{fig:fig1}A).

The precise value of the expected weight, $P_{ij}$, depends upon the particular research question and is flexible to many alternative definitions. The most common definition is the graphical null model, $P_{ij}=\frac{k_i k_j}{2m}$, which gives the expected weight under the null model where each node's strength is preserved exactly but where connections are otherwise formed at random. Here, $k_i=\sum_j W_{ij}$ is node $i$'s strength and $2m=\sum_i k_i$ is the total weight of the network. When a network's connectivity is defined by a correlation matrix, as is the case here, this class of null model may not be appropriate. The connection weights in a correlation matrix represent statistical relationships and are not independent of one another; ``€œrewiring'' the weights of a correlation matrix can result in a randomized matrix that violates these dependencies and may therefore not be mathematically realizable \cite{zalesky2012use}. For this reason, several alternative definitions have been proposed for $P_{ij}$ that are appropriate for use with correlation matrices \citep{macmahon2013community, bazzi2014community}. One such method is the uniform null model, where $P_{ij}=\langle W_{ij}\rangle$. Here, $\langle W_{ij}\rangle$ denotes the average over all pairwise correlation coefficients. Implicitly, then, the uniform null model considers a community to be of high quality if its nodes are more correlated with one another than would be expected given the average correlation of the entire network.

\subsection*{Multi-scale modularity maximization}
Maximizing modularity, $Q$, returns an estimate of a network's community structure. The size and number of these communities defines the \textit{scale} at which a network's community structure is being described. However, the community structure of certain real-world networks may span multiple scales or hierarchical levels, in which case any single-scale community estimate would, at best, miss out on this richness and present an incomplete picture of a network's communities. At worst, the communities returned could be misleading \citep{fortunato2007resolution,lancichinetti2011limits}. In order to detect communities at different scales, the modularity function can be modified by including a tunable resolution parameter \citep{reichardt2006statistical}. Changing the value of this parameter one can effectively shift the scale at which communities are detected, making it possible to uncover communities of different sizes (Figure \ref{fig:fig1}B,C). In the present study, we incorporate the resolution parameter in the following way. Rather than set $P_{ij}=\langle W_{ij} \rangle$, we let $P_{ij}=\gamma \langle W_{ij} \rangle$, where $\gamma$ is the ``€œstructural resolution parameter''€. By absorbing the constant $\langle W_{ij} \rangle$ into the resolution parameter, we can write $P_{ij}=\gamma$, which is similar to the so-called ``constant Potts model'' \citep{traag2011narrow}. Thus, when the value of $\gamma$ is small, many elements of $W_{ij}$ will exceed $\gamma$. At that scale partitions that come close to maximizing $Q(\gamma)$ will produce relatively large communities. On the other hand, when $\gamma$ is large, very few elements of $W_{ij}$ will exceed $\gamma$ and the resulting partitions will feature more communities but contain fewer nodes.

\subsection*{Multi-layer, multi-scale modularity maximization}
A further modification of the modularity function makes it compatible with multi-layer networks \citep{mucha2010community}. A multi-layer network refers to a network whose nodes are linked across different layers \citep{kivela2014multilayer}. Layers may correspond to different connection modalities (e.g. cities connected by air, train, and road travel) or observations of the same network at different instants (e.g. brain networks constructed at different points in a scan session). In the present study, we define multi-layer networks where each layer is the functional connectivity matrix of a different age group. Multi-layer modularity maximization provides a generalization of the traditional single- and multi-scale modularity maximization frameworks, making it useful for handling this type of structure.

Here we briefly discuss the mechanics of multi-layer modularity maximization. Recall that in single-layer modularity maximization the aim was to choose communities so that connections that fall within communities are mostly the positive elements of the modularity matrix, $\mathbf{B}$. Multi-layer modularity maximization works similarly. Let $\mathbf{B}_r=\mathbf{W}_r-\mathbf{P}_r$ be the modularity matrix for layer $r$. We can define a new matrix:

$$
\mathcal{B} = 
\begin{bmatrix}
\mathbf{B}_1 & \ldots & \omega \mathbf{I} \\
\vdots & \ddots & \vdots \\
\omega \mathbf{I} & \ldots & \mathbf{B}_K \\
\end{bmatrix}
$$

The matrix $\mathcal{B}$ has dimensions $[n \times K, n \times K]$, where $n$ is the number of nodes in a single layer and $K$ is the total number of layers. The diagonal of $\mathcal{B}$ contains the single-layer modularity matrices for each of the $K$ layers and each off-diagonal block contains the matrix, $\omega \mathbf{I}$, which is the identity matrix whose diagonal elements are equal to the inter-layer coupling parameter, $\omega$. Multi-layer modularity maximization, then, tries to choose communities so that as many positive elements of $\mathcal{B}$ fall within communities. The associated modularity function is:

\begin{equation}
Q_{multi} = \sum_{ijsr} [ (W_{ijs} - \gamma P_{ijs}) \delta (g_{is}, g_{js}) + \delta (i,j) \cdot \omega] \delta (g_{is}, g_{jr})
\end{equation}

where the community assignment of node $i$ in layer $r$ is given by $g_{ir}$. In the case of the constant null model, we replace $\gamma P_{ijs}$ with $\gamma$. In addition to the resolution parameter, $\gamma$, multi-layer modularity depends upon the value of the inter-layer coupling parameter, $\omega$. When $\omega=0$, nodes are uncoupled across layers and maximizing $Q_{multi}$ is equivalent to maximizing the modularity of each layer independently. When $\omega > 0$, nodes become coupled and can appear in the same community, even across layers. Thus the value of $\omega$ determines the uniformity of community assignments across layers: when $\omega$ is close to zero the community structure of layer $r$ can vary considerably from that of layer $s$; increasing $\omega$ will lead to more homogeneous community structure across slices.

We used a freely available MATLAB software package \citep{jutla2011generalized} to perform multi-layer modularity maximization. This software uses an algorithm similar to so-called ``Louvain'' method of \cite{blondel2008fast} to maximize $Q_{multi}$. Rather than focusing on a single set of parameters, we explored a range of possible values. Specifically, we explored 31 logarithmically spaced values of $\gamma \in [10^{-2},\dots,10^0]$ and $\omega \in [10^{-3},\dots,10^0]$, resulting in $31 \times 31=961$ total parameter combinations. For each pair of parameters, we maximized $Q_{multi}$ once for each multi-layer network in the ensemble of networks (a total of 500 runs). We focused on this partition ensemble and characterized its statistical properties rather than treat any single run as representative.

\subsection*{Network measurements}
Maximizing $Q_{multi}$ returns an ensemble of multilayer partitions. From these partitions we made several measurements.
\begin{enumerate}

\item \textit{Single-layer modularity}: For a single layer $r$ associated with connectivity matrix $\mathbf{W}_r$, we calculated the single-layer modularity: $Q_r (\gamma)= \frac{1}{2m_r}\sum_{ij} [W_{ijr} - \gamma] \delta (g_{ir},g_{jr})$, where $2m_r=\sum_{ij} |W_{ijr}|$ and $g_{ir}$ was the community to which node $i$ in layer $r$ was assigned.

\item \textit{Node flexibility}: Following \cite{bassett2011dynamic}, we calculated a flexibility score as the fraction of all partitions in which node $i$'s community assignment changed from layer $r$ to $s$, which we denote as $f_{ir}$. We also calculated the average flexibility of each layer as $f_r =\sum_i f_{ir}$. We contextualized these scores by comparing them against a permutation-based null model (see Methods, Null models) and expressed them as z-scores, $z_{ir}$ and $z_{r}$. The z-scores indicate how much more or less flexible nodes or layers were than chance.

\item \textit{Association matrix}: We also calculated the association matrix $\mathbf{T}=[T_{ij}]$, where $T_{ij} = \frac{1}{K}\sum_r \delta(g_{ir}, g_{jr})$ for each partition in the ensemble. Each element of the association matrix counts the fraction of layers in which nodes $i$ and $j$ were assigned to the same community. We expressed the association matrix as the average across all partitions in the partition ensemble.

\item \textit{Participation coefficient}: Given a partition, one can calculate how a node's connections are distributed across modules using the participation coefficient \citep{guimera2005functional}: $p_i = 1 - \sum_c (\frac{\kappa_{ic}}{k_c})^2$, where $\kappa_{ic}$ is the total weight of connections node $i$ makes to module $c$. For a signed network (e.g. a correlation matrix), we calculate the participation coefficient of positive and negative links separately: $p_i^{\pm} = 1 - \sum_c (\frac{\kappa_{ic}^{\pm}}{k_c^{\pm}})^2$.

\end{enumerate}

\subsection*{Null models}
We used two different null models against which we compared the results presented in the main text. To test the robustness of flexibility scores, we constructed null multi-layer partitions as part of the \textit{permutation null model}. Let $\mathbf{G}$ be a multi-layer partition such that $\mathbf{G}=\{ \mathbf{g}_1,\ldots,\mathbf{g}_K \}$, where $\mathbf{g}_r=\{g_{1r},\ldots,g_{nr}\}$ is the partition of nodes in layer $r$. In other words, $\mathbf{g}_r$ maps node $i$ in layer $r$ to one of $C$ communities. The permutation null model leaves these node-level mappings intact, but permutes the order of the layers. For example, if $\mathbf{G}=\{ \mathbf{g}_1,\mathbf{g}_2,\mathbf{g}_3,\mathbf{g}_4 \}$, then a partition generated by the permutation null model might look like $\mathbf{G}'=\{ \mathbf{g}_3,\mathbf{g}_1,\mathbf{g}_4,\mathbf{g}_2 \}$. We used this model to test the null hypothesis that, given the observed multi-layer partitions, the flexibility scores we obtained could be explained by a reordering of the single-layer partitions.

The second null model against which we compared our results was the \textit{network null model}. This model involved constructing null multi-layer networks, $\mathcal{W}'=\{ \mathbf{W}_1',\ldots,\mathbf{W}_K' \}$. Whereas $\mathbf{W}_r$ was a composite matrix representative of a subset of participants of roughly the same age, $\mathbf{W}_r'$ was a composite of randomly selected subjects. The number of subjects used to construct $\mathbf{W}_r'$ was exactly equal to that of $\mathbf{W}_r.$ We constructed 500 realizations of $\mathcal{W}'$ and optimized their multi-layer modularity using precisely the same approach as applied to the empirical multi-layer networks. The output of these procedures, then, was used to test the dependence of our results on the age composition of the multilayer networks. These models were used in two instances: first, we tested whether the correlation magnitude of age and single-layer modularity, $Q_r(\gamma)$, could have been obtained by chance; secondly, we tested whether the flexibility of partitions obtained from $\mathcal{W}'$ was comparable to the flexibility of partitions obtained from $\mathcal{W}$.

\section*{Results}
\begin{figure}[ht]
\centering
\includegraphics[width=\linewidth]{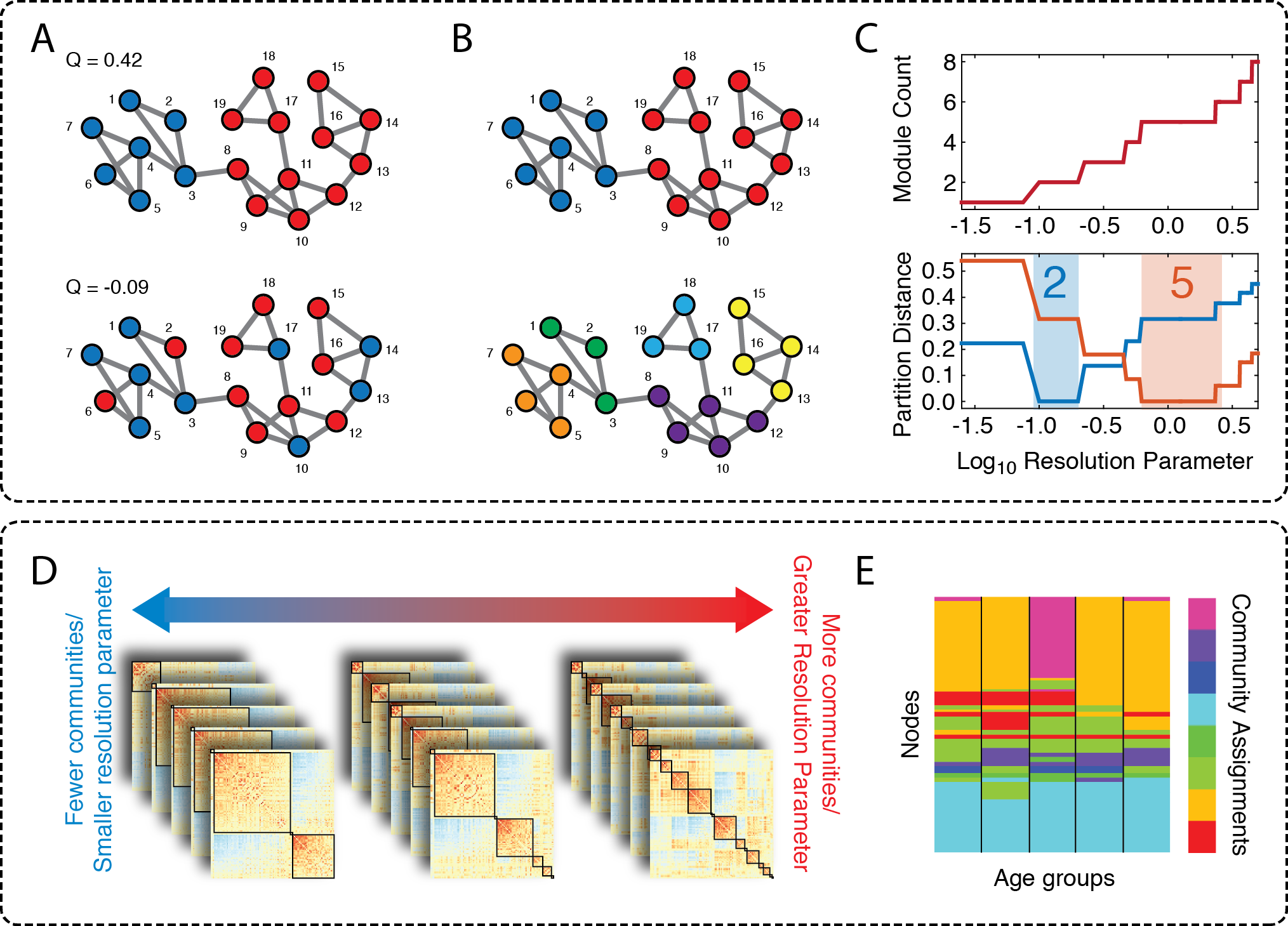}
\caption{Modularity maximization schematic. A) Example toy network divided into two communities according to two different partitions. The top and bottom partitions correspond to modularity scores of $Q=0.42$ and $Q=-0.09$, respectively. Accordingly, the top partition would be regarded as superior. B) Single-scale modularity maximization overlooks networks with multi-scale/hierarchical community structure. Here, we highlight two potential partitions of the same network into two (top) and five (bottom)  communities. C) To detect these and other potentially interesting partitions, we use \textit{multi-scale modularity maximization}. This process entail adding a resolution parameter, $\gamma$, to the modularity equation. In this toy example we maximized the modularity: $Q(\gamma) = \sum_{ij} [A_{ij} - \gamma P_{ij}]\delta(g_i,g_j)$, where $P_{ij} = \frac{k_i k_j}{2m}$ with $k_i = \sum_j A_{ij}$ and $2m = \sum_i k_i$. Over the range $\gamma \in [10^{-1.5}, 10^{0.5}]$ we find ranges of $\gamma$ where we uncover the two- and five-community partitions exactly. To measure the similarity of detected partitions with the planted partitions we used, as a partition distance, variation of information. D) The general strategy of this paper was to divide the $N=316$ NKI-Rockland participants into $K$ equally-sized groups according to their age. For each group we construct a representative functional connectivity matrix, which we submit to a multi-layer, multi-scale modularity maximization algorithm. By varying the resolution parameter we track community structure across the lifespan over a range of organization scales. E) For a given resolution parameter, the community detection algorithm partitions nodes each age group into a single community.}
\label{fig:fig1}
\end{figure}

The aim of this study was to characterize age-related changes in the community structure of functional brain networks at multiple scales (Figure \ref{fig:fig1}D-E). We constructed representative connectivity matrices, $\mathbf{W}=[W_{ij}]$, for $K$ different age groups. These matrices were then arranged to form multi-layer networks, $\mathcal{W}=\{\mathbf{W}_1, \dots ,\mathbf{W}_K\}$, with each age group represented as a layer (Figure \ref{fig:fig1}D). Using a resampling procedure, this process was repeated 500 times, thereby generating 500 estimates of $\mathcal{W}$. The analyses described herein were carried out over this ensemble of multi-layer networks with $K=5$ age groups. The resulting age ranges for each group were 8.3-22.4, 22.4-41.0, 41.0-51.4, 51.4-62.8, and 62.8-83.4 years. For each multi-layer network in the ensemble we maximized a multi-layer modularity function in order to obtain community assignments for brain regions across layers. This procedure allowed us to track the formation, evolution, and dissolution of communities with age. The modularity maximization process was dependent upon two parameters, $\gamma$ and $\omega$, sometimes referred to as the ``€œstructural'' and ``€œinterlayer''€ resolution parameters, respectively. By tuning these parameters we were able to examine communities of different size and number (Figure \ref{fig:fig2}A-C).

\subsection*{Multi-layer modularity maximization uncovers known ICNs}
\begin{figure}[ht]
\centering
\includegraphics[width=\linewidth]{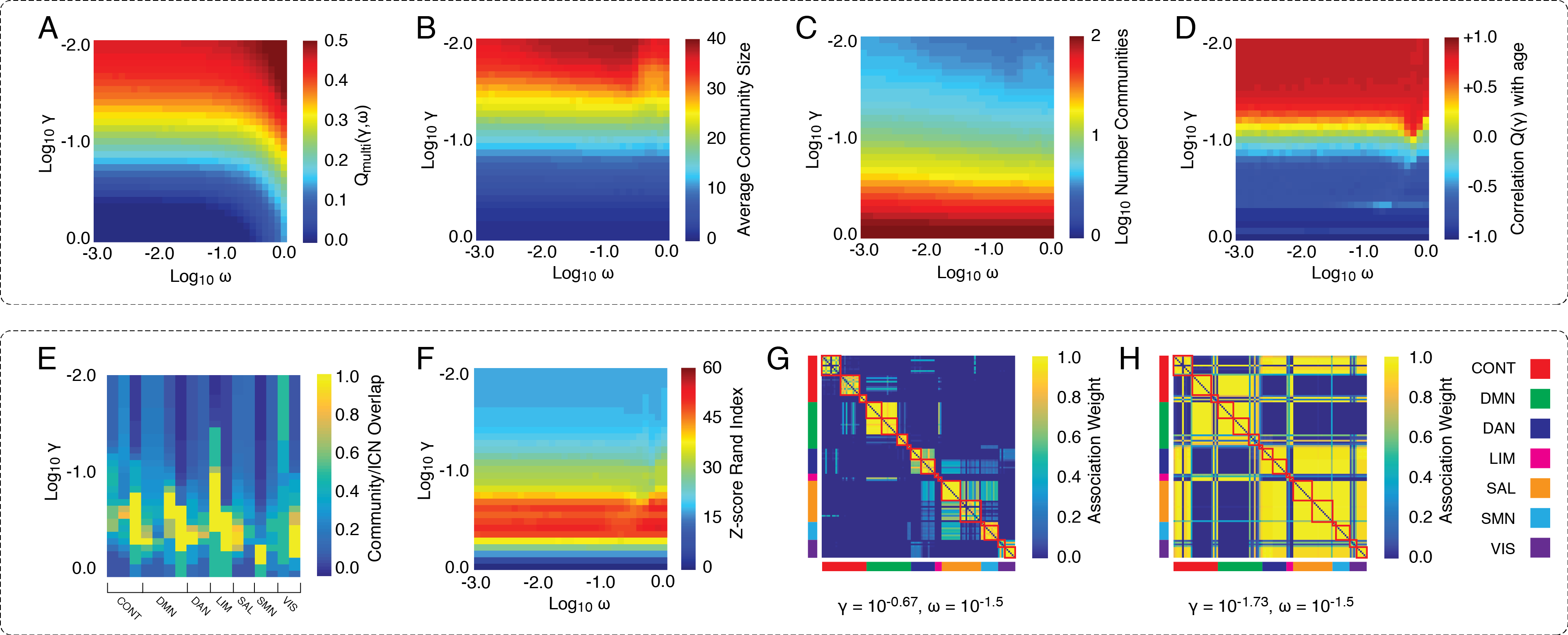}
\caption{Statistics of multi-layer, multi-scale modularity maximization as a function of resolution parameters, $\gamma$ and $\omega$. A) Multi-layer modularity, $Q_{multi}(\gamma, \omega)$. B) Average size of single-layer communities. C) Log-transformed number of communities. D) Correlation of single-layer modularity with age, $\hat{r}_{age,Q_r (\gamma)}$. Note that as $\gamma$ increases the correlation coefficient decreases more or less monotonically. E) For each of the seventeen ICNs defined by \cite{yeo2011organization} we show the mean similarity of that ICN to the best-matched communities detected by the multi-layer, multi-scale modularity maximization. F) Similarity (z-score of Rand index) of detected partitions to the entire Yeo ICN partition. G-H) Association matrix, $\mathbf{T}$, which measures the fraction of all partitions in which two nodes were assigned to the same community. The rows and columns of the association matrices are ordered according to the Yeo ICNs.}
\label{fig:fig2}
\end{figure}

We first tested whether the communities uncovered with multi-layer modularity maximization were similar to those reported in an earlier large-scale study \citep{yeo2011organization}. In that study, the cerebral cortex was clustered into seventeen ICNs. We compared detected multi-layer communities to the ICN partition in two ways. In both cases we decomposed each multi-layer partition into a set of $K$ single-layer partitions. We first calculated the similarity of each community in the single-layer partitions with the ICN it most closely resembled. We used the Jaccard index as a measure of similarity. For two sets $X=\{x_1, \ldots, x_m \}$ and $Y=\{y_1, \ldots, y_n \}$, their Jaccard similarity is defined as:

\begin{equation}
J_{XY} = \frac{|X \cap Y|}{|X \cup Y|}
\end{equation}

where $J_{XY}$ is bounded by the interval $[0,1]$ and where a value of 1 indicates that sets $X$ and $Y$ perfectly overlap. For each ICN we defined  $X_{ICN} = \{i \in ICN \}$ as the set of all nodes assigned to that ICN. Similarly, for community $g$ in any single-layer partition, we defined $X_g = \{i \in g \}$. From these two sets we calculated the similarity of any community $g$ with any $ICN$ as $J_{X_{ICN},X_g}$. Furthermore, within each single-layer partition we identified the community that was maximally similar to each of the seventeen ICNs and averaged this maximum similarity across all single-layer partitions. We repeated this process for all single-layer partitions at each value of $\gamma$, which allowed us to identify the scale at which each ICN was most reliably detected as a community. We find that for many ICNs there exists a value of $\gamma$ at which it is exactly recovered as a community (i.e. $J_{X_{ICN},X_g} = 1$) (Figure \ref{fig:fig2}E). Interestingly, we observed that the peak similarity for most ICNs occurred within the range $10^{-1} < \gamma < 10^{0}$, suggesting that there was a range at which the detected communities were, on average, highly similar to the Yeo ICNs.

To better understand this relationship, we calculated the similarity of each single-layer partition to the ICN partition as the z-score of the Rand index \citep{traud2011comparing}. For each set of parameters, $\{ \gamma, \omega \}$, we calculated the mean similarity over all single-layer partitions (Figure \ref{fig:fig2}F). We found that the z-score Rand index peaked within a range similar that of the ICN-level Jaccard index. Upon further examination of the single-layer partitions within this range, we found that many of the communities simultaneously matched those observed in the ICN partition. To help visualize this correspondence, we fixed $\gamma = 10^{-0.67} \approx 0.214$ and $\omega = 10^{-1.50} \approx 0.032$ and constructed the association matrix, $\mathbf{T}$, whose elements $T_{ij}$ were equal to the fraction of times that nodes $i$ and $j$ were assigned to the same community in any layer across the partition ensemble. We then reordered the rows and columns of $\mathbf{T}$ so that nodes belonging to the same ICN appeared next to each other (Figure \ref{fig:fig2}G). The block diagonal structure of the matrix indicates that nodes assigned to the same ICN in \cite{yeo2011organization} were also usually assigned to the same algorithmically-detected community in our study. For completeness, we also show an association matrix constructed from partitions at a scale that was not especially similar to the ICN partitions ($\gamma = 10^{-1.73} \approx 0.019$).

\subsection*{Age-dependent changes in community structure are scale dependent.}
Next, we explored the modularity of functional brain networks across multiple scales. In order to assess age-related changes in modularity, we decomposed multi-layer partitions into single-layer partitions, as described earlier, and calculated for each layer, $r$, its single-layer modularity score, $Q_r(\gamma)$, which provides an estimate of the extent to which communities in layer $r$ are well-defined and segregated from one another. We found that $Q_r(\gamma)$ varied systematically with age, though whether it increased or decreased depended on the value of $\gamma$ (i.e. the scale of communities). While our community detection algorithm allowed us to investigate community structure over a much broader range of scales, we choose to focus on a more manageable, though still representative, subset of scales: A ``coarse'' scale ($\gamma = 10^{-1.73} \approx 0.0185$) at which the network was divided into a small number of communities ($3.2\pm0.5$ communities per layer) and a ``fine'' scale ($\gamma = 10^{-0.67} \approx 0.22$) which resulted in divisions of the network into many small communities ($15.4\pm1.0$ communities per layer). In both cases we set $\omega = 10^{-1.5} \approx 0.032$. This value is the median $\omega$ that we examined, and represents a parameter value where communities are variable from layer to layer but where we still find communities that persist across \textit{all} layers.

\begin{figure}[ht]
\centering
\includegraphics[width=\linewidth]{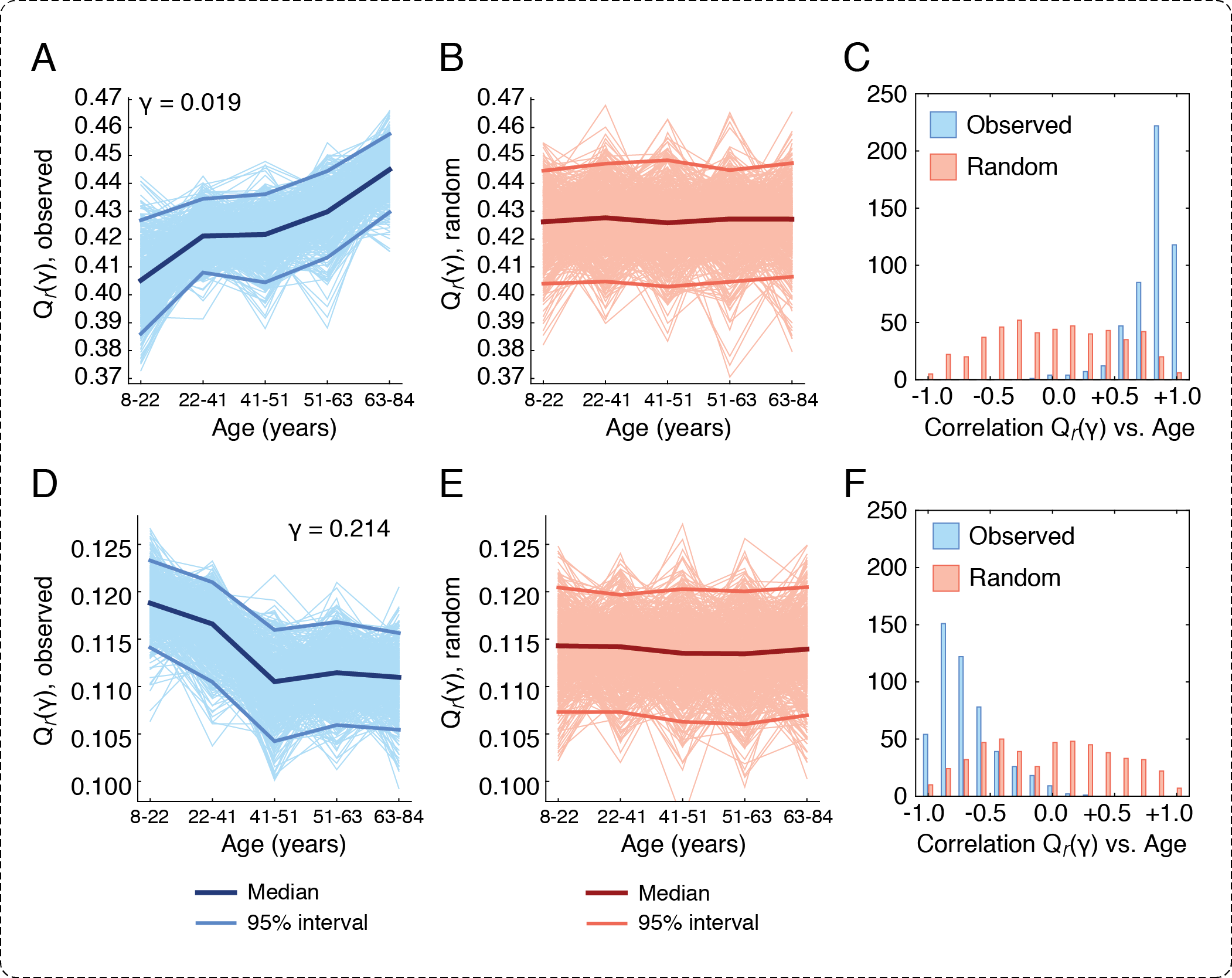}
\caption{Scale-dependent relationship of single-scale modularity, $Q_r(\gamma)$, with age. A) We calculated the single-layer modularity score for each layer based on the partition ensemble generated using multi-layer, multi-scale modularity maximization. Here we plot each of those curves as a function of age when we fix the parameters $\gamma = 0.019$ and $\omega = 0.032$. B) We show a similar plot to that of Panel A but where the modularity maximization algorithm was carried out on random networks. C) We compare the distribution of correlation coefficients, $\hat{r}_{age,Q_r(\gamma)}$, for both observed and randomized cases. Panels D-F recapitulate those of A-C but with $\gamma = 0.214$.}
\label{fig:fig3}
\end{figure}

At both coarse and fine scales we calculated $\hat{r}_{age,Q_r(\gamma)}$, which gives the magnitude to which $Q_r(\gamma)$ and age are correlated with one another. At the coarse scale, we found that $Q_r(\gamma)$ increased with age (Figure \ref{fig:fig3}A), suggesting that large communities become more segregated across the lifespan (median correlation of $\hat{r} = +0.84$ and inter-quartile range of $[+0.73,+0.92]$). We also found that the observed correlation coefficients were statistically stronger (more positive) than those obtained under a network null model (t-test, $df=998$, $t=32.37$, $p \approx0$) (Figure \ref{fig:fig3}B,C). Conversely, at the fine scale, we found that $Q_r(\gamma)$ decreased with age (Figure \ref{fig:fig3}D) (median correlation of $\hat{r}=-0.75$ and inter-quartile range of $[-0.88,-0.58]$ (Figure \ref{fig:fig3}E,F). In this case the observed correlation coefficients were statistically more negative than those obtained under a network null model (t-test, $df=998$, $t=-27.46$, $p\approx0$). We repeated these analyses using different numbers of age groups ($K={4,6,7}$; Figure \ref{fig:figs2}A-I), a different cortical parcellation \citep{destrieux2010automatic} (Figure \ref{fig:figs2}), and after applying an additional motion-correction step (regressing frame-wise displacement from single-layer modularity scores) (Figure \ref{fig:figs3}), and found that in all cases there was evidence in support of the hypothesis that $Q_r(\gamma)$ follows scale-specific trajectories.
of scale-dependent trajectories of $Q_r(\gamma)$ with age.

\begin{figure}[ht]
\centering
\includegraphics[width=\linewidth]{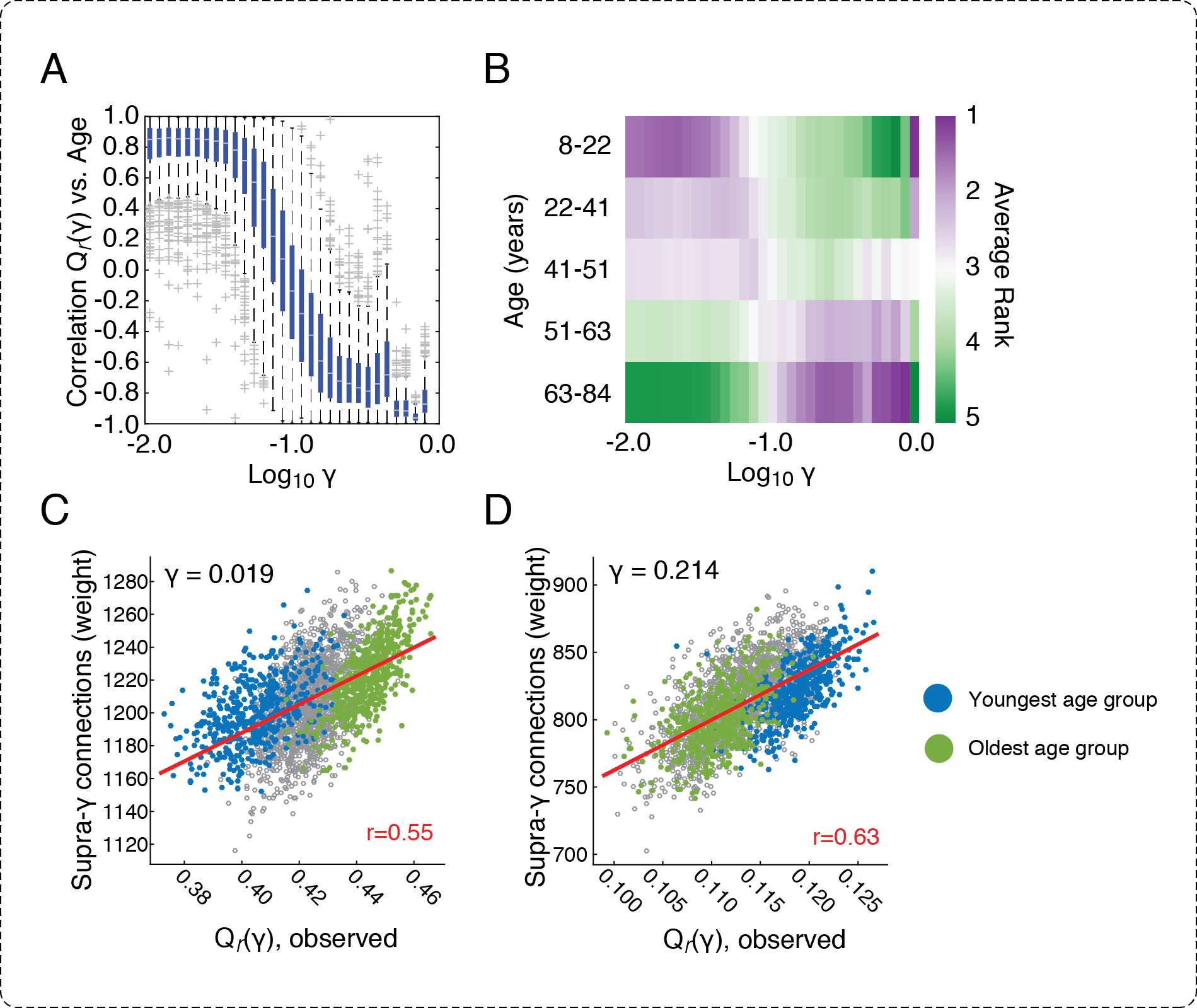}
\caption{Explanation for the scale specificity of $\hat{r}_{age,Q_r(\gamma)}$. A) Box and whisker plot displaying the distribution of correlation coefficients for all values of $\gamma$; importantly, there is a monotonic decrease in the median correlation as $\gamma$ increases. To explain this shift from positive to negative correlations, we calculated the total weight of connections above that exceed the value of the resolution parameter, $\gamma$, which are the only class of connections capable to contributing positive modularity. B) We rank each layer in the multi-layer network ensemble according to their total supra-$\gamma$ weight. We find that this ranking closely mirrors the shape of $\hat{r}_{age,Q_r(\gamma)}$ as a function of $\gamma$. C) We plot the total weight of connections in excess of $\gamma = 0.019$. At this resolution, the oldest age group has a greater total weight of supra-$\gamma$ connections than the younger age group and, as a consequence, tend to achieve greater modularity scores. D) We repeat the same analysis but with $\gamma = 0.214$, a resolution at which the youngest age group now tends to have a greater total weight of supra-$\gamma$ connections than the oldest and, similarly, greater modularity scores.}
\label{fig:fig4}
\end{figure}

We found that as $\gamma$ increased, $\hat{r}_{age,Q_r(\gamma)}$ decreased more or less monotonically (Figure \ref{fig:fig4}A). In order to uncover the origin of this apparent interaction of modularity and age with scale, we examined the distribution of connection weights in each layer. The modularity of any layer, $Q_r(\gamma)$, can only receive positive contributions from connections that exceed the weight expected under some null model (here, the expected weight was the same for all connections and was equal to $\gamma$). More specifically, if layer $r$ contains many connections that satisfy the condition $W_{ijr}-\gamma>0$, then that layer has the capacity to achieve a large modularity score, $Q_r(\gamma)$ provided that the supra-$\gamma$ connections cluster within communities. We examined each layer (age group) at every value of $\gamma$ and calculated the total weight of connections that exceeded $\gamma$. We found that when $\gamma \approx 0$, the oldest age group contained the greatest number of supra-$\gamma$ connections while the youngest age group contained the fewest. As $\gamma$ was increased, however, this relationship reversed (Figure \ref{fig:fig4}B-D) and the youngest age group exhibited the greatest number of supra-$\gamma$ connections. These results suggest that whether $Q_r(\gamma)$ increased or decreased with age was a consequence of the shape of the connection weight distribution and its relation to the resolution parameter, $\gamma$.

\begin{figure}[ht]
\centering
\includegraphics{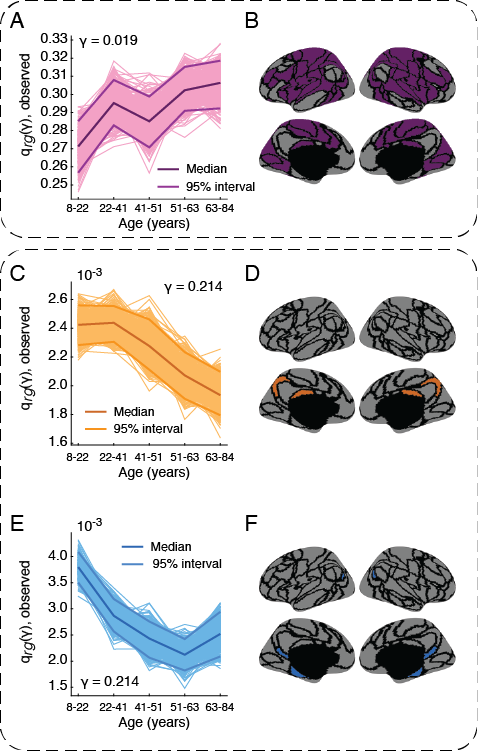}
\caption{Community contributions to lifespan changes in modularity. A) When $\gamma = 0.019$ a single community composed of task-positive regions is most responsible for the increase in $Q_r(\gamma)$. When $\gamma = 0.214$ two communities drive the changes in modularity: B) A community comprised of posterior cingulate and precuneus; C) A community comprised of retrosplenial and parahippocampal cortex along with inferior parietal lobule.}
\label{fig:fig5}
\end{figure}

Finally, we wanted to determine which communities were most responsible for driving the age-related increases and decreases in $Q_r(\gamma)$. To this end, we obtained consensus communities (Figure \ref{fig:figs4}) for both coarse and fine scales and calculated the modularity contribution, $q_{rg}$, that each consensus community, $g$, made to the total single-layer modularity $Q_r(\gamma)$ (these measures are related to one another by $Q_r(\gamma)=\sum_g q_{rq}$ where $q_{rg} = \sum_{ij\in g}[W_{ijr} - \gamma]$). We found that at a coarse scale, a single community accounted for 70\% of the total modularity and that this community's modularity was positively correlated with age, suggesting that it was the primary driver of the age-dependent evolution of the single-layer modularity score (Figure \ref{fig:fig5}A,B). This community was spatially distributed and aligned closely with the brain's task-positive system. At finer scales, the larger communities fragmented into smaller communities whose modularity displayed distinct age-related trajectories. Several communities exhibited decreased modularity, including two communities that both contributed positive modularity ($q_{rg}>0$) in the youngest age groups but went on to contribute zero or negative modularity ($q_{rg}<0$) with increased age. The first community was comprised of portions of the posterior cingulate and precuneus reported in \cite{yeo2011organization} as part of the control network (Figure \ref{fig:fig5}C,D), but more often associated with the default mode network as hub or core regions \citep{fransson2008precuneus, utevsky2014precuneus}, while the second community was comprised of retrosplenial and parahippocampal cortex and parts of the intraparietal lobule associated with the default mode network \citep{buckner2004memory} (Figure \ref{fig:fig5}E,F).

\subsection*{Community structure varies with age}
Another important aim of this paper was to quantify the extent to which brain regions' community assignments changed with age. The inter-layer resolution parameter, $\omega$, played an important role in this regard. When $\omega = 0$ communities do not span layers; i.e. communities in layer $r$ will not appear in any other layer. However, when $\omega > 0$, nodes become coupled to one another across layers and communities in one layer can appear in others. In this section we chose not to focus on a single $\omega$ value, demonstrating the robustness of our results by reporting a range of values.

\begin{figure}[ht]
\centering
\includegraphics[width=\linewidth]{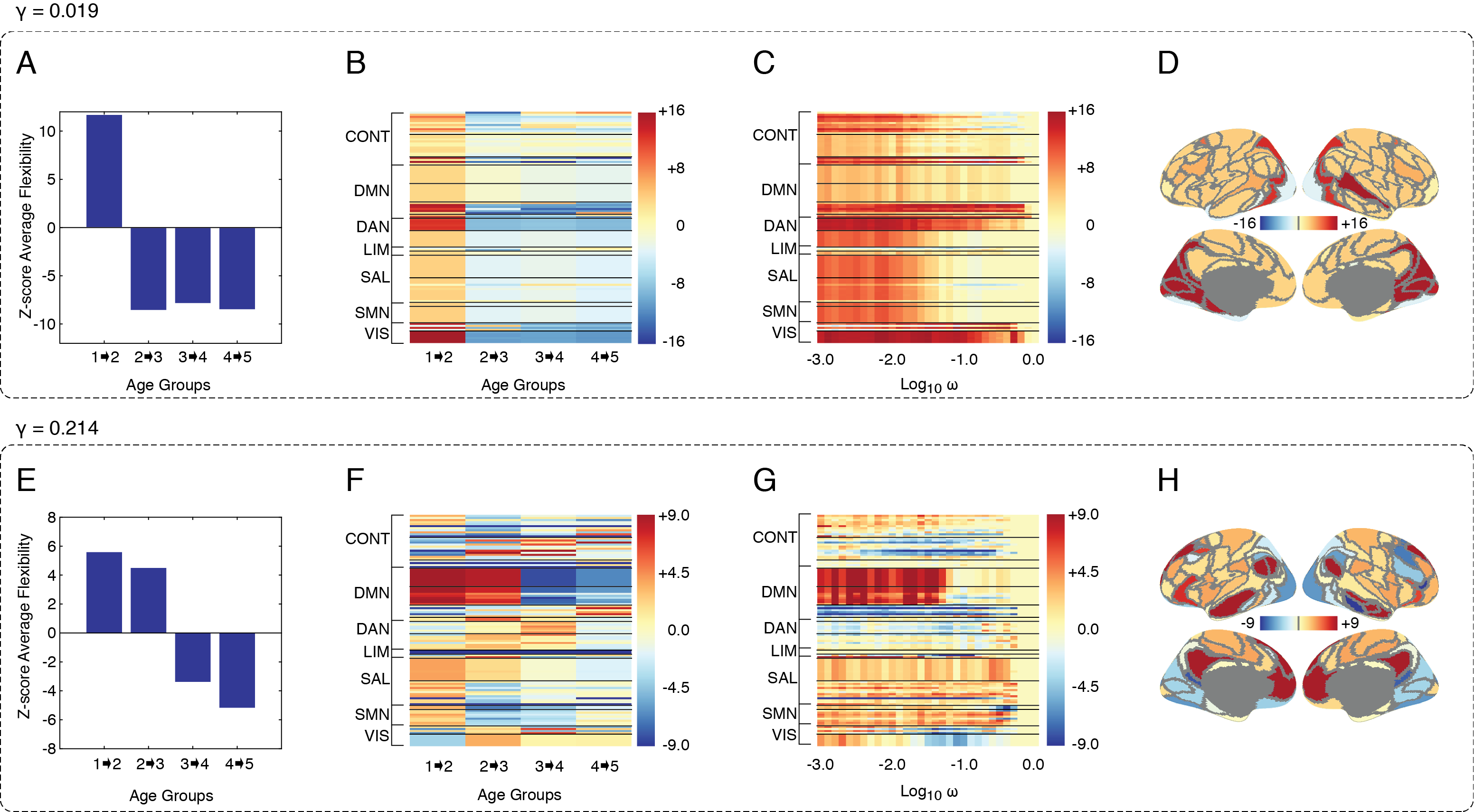}
\caption{Flexibility profiles at coarse ($\gamma = 0.019$) and fine ($\gamma = 0.214$) scales. A) Z-score of average flexibility between each sequential age group. Positive z-scores indicate greater-than-expected flexibility (variable community structure). B) Z-scores of each region's flexibility between sequential age groups. C) A depiction of the flexibility between age groups 1 and 2 across the full range of $\omega$ values. D) A topographic depiction of the flexibility between age groups 1 and 2 with $\omega = 10^{-1.5} \approx 0.032$. Panels E-H feature the same information as A-D but with $\gamma = 0.214$.}
\label{fig:fig6}
\end{figure}

In order to determine whether brain regions change communities with age, we calculated the standardized flexibility, $z_{ir}$, of each region, which indicated the number of times that node $i$ changed its community assignment from layer $r$ to $r+1$ across the partition ensemble. From the node-level flexibility scores we also calculated the standardized average flexibility of each layer, $z_r$. We found that average flexibility was consistently greatest between the first (8.3-22.4 years) and second (22.4-41.0 years) age groups, while flexibility was near or below chance levels for all other age groups (Figures \ref{fig:fig6}A,E).

We also examined the flexibility profiles of individual brain regions. As expected, individual nodes were also most flexible between the first and second layers (Figures \ref{fig:fig6}B,F). From layer one to two and at coarse scales, we found that brain regions associated with control (dorsal precuneus and dorsal pre-frontal cortex), default mode (parahippocampal and retrosplenial cortex), dorsal attention (superior parietal lobule, parieto-occipital, and temporo-occipital cortex), and visual systems (striate and extra-striate cortex) were most flexible (Figures \ref{fig:fig6}C,D). At finer scales, the flexibility pattern was different; the most flexible regions were associated almost exclusively with the default mode network (parts of temporal, posterior cingulate, and both dorsal and medial pre-frontal cortex, along with the inferior parietal lobule) (Figures \ref{fig:fig6}G,H). At this scale, a small number of regions were far less flexible than expected, including retrosplenial cortex. We explored the flexibility using an alternative null model (Figure \ref{fig:figs5}) and for different numbers of age groups (Figure \ref{fig:figs6}). These additional analyses provided additional evidence for scale-dependent changes in community across the lifespan. In general, they all agree that at coarse scales, the greatest flexibility occurs early in life; at finer scales the results occasionally diverge.

%reproduced the results in this section using an alternative null model \textcolor{red}{Figure S9} and for different numbers of age groups \textcolor{red}{Figures S4-S6}. Increasing the number of age groups allowed us to characterize, more precisely, when the brain's community structure is most flexible. Indeed, with a greater number of age groups, we found that the flexibility between the first two age groups approached chance levels. Instead, community structure appeared most flexible between age groups two (22.4-40.9 years) and three (40.9-51.4 years) for $K=6$ \textcolor{red}{Figure S5} and layers three (30.3-47.3 years) and four (47.3-54.7) for $K=7$ \textcolor{red}{Figure S6}.

\subsection*{Functional roles change with age}
A final focus of our study was to characterize brain regions' functional roles with respect to modules, which we assessed using the participation coefficient \citep{guimera2005functional}. The participation coefficient measures how uniformly distributed a node's connections are across modules, with values close to one indicating greater uniformity (See Methods). The participation coefficient depends not only on the distribution of a node's connections, but also on the network's modular structure, which makes it difficult to disentangle the effect of one from the other. For this reason, we restricted our analysis to partitions obtained with $\omega = 1$, for which single-layer community structure was consistent across all layers. Doing so allowed us to attribute any age-related changes in nodes' participation coefficients to alterations in the distribution of nodes' connections rather than fluctuations in community structure. We also restricted our analysis to the participation coefficient of positive connections, though an analogous score can be calculated for negative connections \citep{rubinov2011weight}.

To identify regions whose participation coefficients changed with age, we calculated the Pearson's correlation of each region's participation coefficient across layers (age). We repeated this for each partition in the ensemble, which generated a distribution of correlation coefficients. We focused on regions with distributions whose interquartile range excluded the value of zero and, for these regions, calculated the mean change in participation coefficient from the first to the final layer.

\begin{figure}[ht]
\centering
\includegraphics[width=\linewidth]{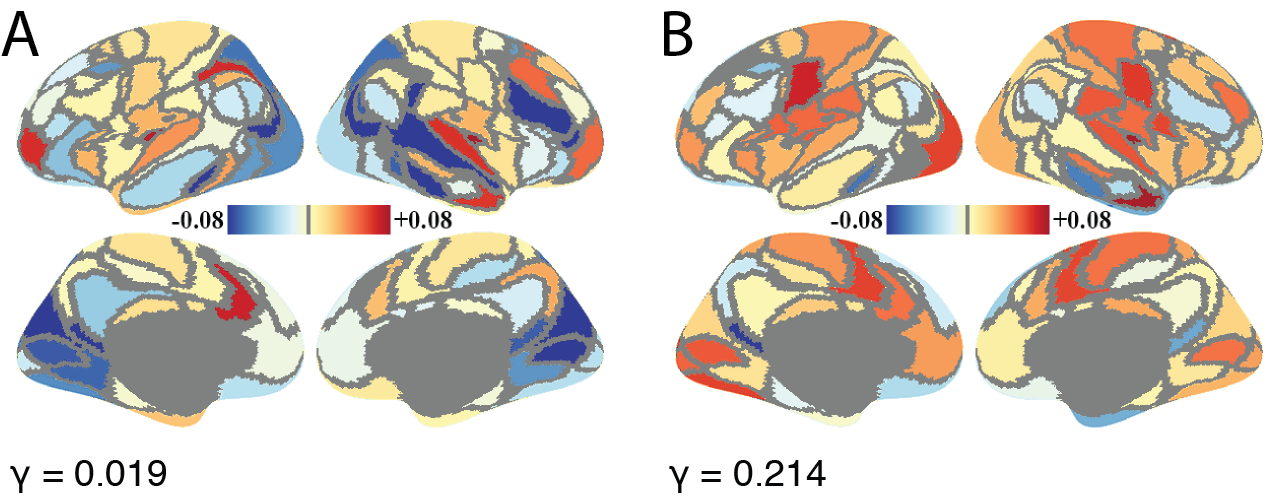}
\caption{Changes in participation coefficient from the youngest age group to the oldest age group. A) At the coarse scale with $\gamma = 0.019$ and $\omega = 0.032$. B) At the fine scale with $\gamma = 0.214$ and $\omega = 0.032.$}
\label{fig:fig7}
\end{figure}

\begin{figure}[ht]
\centering
\includegraphics[width=\linewidth]{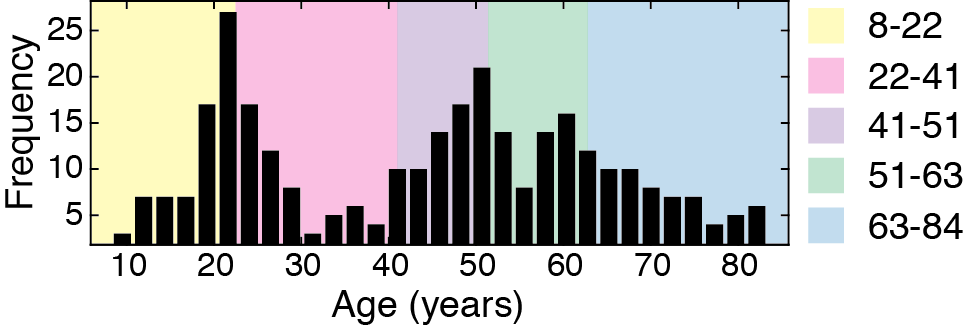}
\caption{Age distribution for all $N=316$ NKI-Rockland participants analyzed in the present study. The colors behind the histogram indicate the boundaries of the age groups ($K=5$), which were determined so that each group contained an equal number of participants.}
\label{fig:figs1}
\end{figure}

At coarse scales the regions whose participation coefficient increased most consistently and by the greatest amount were portions of the insula associated with the somato-motor network, pre-frontal regions associated with saliency/ventral attention and control networks, and temporal regions associated with control and default mode networks (Figure \ref{fig:fig7}A). Other regions consistently decreased their participation coefficient, including parietal-occipital, retrosplenial, and striate/extra-striate cortices in dorsal attention, default mode, and visual networks, respectively.

At fine scales a different pattern of change emerged. The participation coefficient of regions associated with the somatomotor network continue to increase but are joined by striate/extra-striate, posterior cingulate, and medial-frontal cortex in the visual, control and saliency/ventral attention networks, respectively (Figure \ref{fig:fig7}B). At this scale, the regions that exhibit the biggest decreases in participation coefficient are associated with the default mode network and include retrosplenial cortex along with lateral/dorsal pre-frontal cortex and inferior parietal lobule as well as control regions in temporal cortex.

\section*{Discussion}
This study describes the multi-scale evolution of communities in the brain's functional connectivity across a large part of the human lifespan. We demonstrated that multi-layer/multi-scale community detection delivers communities that are highly consistent with known ICNs. We then show that the evolution of communities with age cannot be fully characterized at a single scale. Rather, we found that communities of different sizes and compositions allow us to uncover different (though complementary) descriptions of age-related change. At a coarse scale, we found that community structure becomes more modular and less functionally integrated with age. At fine scales this relationship reversed, and communities became less segregated. To determine which regions change their affiliation with communities and at what point in the lifespan these changes occur, we leveraged the concept of node flexibility. We found that the pattern of change was scale-specific and that most changes in community structure occurred between the first two layers (age ranges of 8.3-22.4 and 22.4-41.0 years, respectively). Finally, we quantified the extent to which a region's connections were distributed across modules using the participation coefficient. We showed that participation coefficients follow age-related trajectories, with somato-motor and retrosplenial cortex forming proportionally more and less positive connections to other modules, respectively.

\subsection*{Age-related change in community structure varies with scale}
Most previous studies of functional communities have characterized their organization at a single scale without explicitly examining community structure at other potentially biologically meaningful scales. Though there have been some efforts to study multi-scale or hierarchical modularity in brain networks, their focus has been on the advancement of theory \citep{betzel2013multi} or methods, e.g. cortical parcellations \citep{doucet2011brain}. Using a multi-scale approach, we recapitulated some important results from the extant literature. In particular, we demonstrated that at fine scales communities grow less segregated with age \citep{meunier2009age, betzel2014changes, chan2014decreased, cao2014topological, geerligs2014brain}. Exemplifying this decreased segregation were two communities that became less modular with age. The first community was comprised of parahippocampal and retrosplenial cortex while the second community contained areas in the posterior cingulate associated with cognitive control. These same regions were also among the least flexible (maintained allegiance to the same community) and exhibited the greatest decrease in participation coefficient (a larger proportion of their connections were made to regions in the same module). Interestingly, the regions comprising the first community have overlapping cognitive-behavioral profiles, and have been implicated in episodic memory, navigation, and orientation \citep{vann2009does}. Of particular relevance to the present study is the relationship of these regions to memory and aging, where disruptions to subnetworks involving the retrosplenial and posterior cingulate cortices have been posited as neurobiological underpinnings of age-related declines in memory \citep{buckner2004memory, sambataro2010age}.

The multi-scale approach also allowed us to examine community structure at a coarse scale typified by few large communities. At this scale we found communities that corresponded closely to a division of the cortex into task-positive/negative systems \citep{golland2008data}. These communities become more segregated with age, a relationship driven by an increase in the modularity (segregation) of task-positive regions. Together with the concurrent decrease in the segregation of communities at finer scales, these results suggest that the brain's task-positive system dissociates from the default mode network and becomes more integrated but in a non-specific way, such that no particular task-positive sub-system is favored. This finding supports the de-differentiation hypothesis wherein brain regions lose the specificity of their functional partners with age \citep{grady2012cognitive}. As a possible consequence, older adults can exhibit broader spatial patterns of activity across task-positive regions compared to performance-matched younger adults, possibly as compensation for declining cognitive ability or due to impaired recruitment mechanisms \citep{cabeza2002aging}.

We also assessed community temporal stability across the lifespan by calculating regions' flexibility scores. We found that flexibility was greatest, on average, early in life, though individual regions exhibited greater-than-expected flexibility across all stages, suggesting that the brain's functional systems undergo continuous refinement. These findings align with theories of the plastic brain \citep{pascual2005plastic}, wherein subjective experience in all stages of life \citep{li2006neuromodulation} promotes cortical reorganization. If we interpret these results from a practical perspective, they suggest that divisions of the cortex into canonical ICNs may not completely characterize the cognitive architecture of individuals that fall outside of the age range of young adults.

\subsection*{Possible mechanisms}
The nature of our data and the structure of our analyses make it difficult to directly identify neurobiological mechanisms that drive changes in community structure with age. There are several possible scenarios. One possibility is that the observed changes in community structure are driven by changes in the underlying anatomy. Across the lifespan, the brain's white and gray matter architecture undergoes continuous developmental refinement \citep{sowell2003mapping, barnea2005white, douaud2014common}. These refinements, which are region-specific and include changes in volume and myelination status, contribute to defining the brain's anatomical network. A substantial amount of variation in the magnitude of functional connectivity can be explained by the pattern in which anatomical connections, reflecting white matter fascicles, are configured \citep{honey2009predicting, hermundstad2013structural, goni2014resting, misic2015cooperative}, and there is evidence that the strength of this relationship varies with age \citep{hagmann2008mapping}. Thus, by influencing functional connectivity patterns, it is possible that age-related changes in anatomical connectivity ultimately underpin the observed variation in functional communities.  The NKI-Rockland lifespan sample includes diffusion imaging scans, which makes it possible to construct anatomical networks for each participant. Future work should investigate further the relationship between these two classes of networks.

\subsection*{Community detection for functional brain networks}
In this study we utilize a set of multi-scale and multi-layer methods for studying brain networks, which provide additional depth to the methods currently being used in the field. The multi-layer approach, for instance, confers obvious advantages, especially in the context of community detection. Most community detection approaches partition the nodes of single-layer networks into communities but leave it up to the user to match the communities detected in one layer to those in another. The multi-layer method used here partitions all layers simultaneously, maintaining a consistent set of community labels across layers and thereby automating the matching process \citep{mucha2010community}. This has implications for studies that examine differences in community structure as the result of experimental manipulation or disease \citep{alexander2012discovery}. A multi-layer approach to community detection makes the comparison of communities between groups straightforward. The multi-layer approach also makes it easier to analyze networks whose layers are ordinally related to one another (e.g. layers that correspond to particular ages or time points). As noted in earlier studies \citep{cole2014intrinsic, bassett2015learning}, nodes' community assignments can be tracked across layers, making it possible to quantify the instant at which a node moves to a new community, or to find frustrated nodes with no consistent community assignment.

The present study proposes several methodological innovations. First we introduce a sub-sampling procedure for constructing composite brain networks. Because resting-state scans are of finite length and may thus provide an incomplete sample of the brain's ``dynamic repertoire'' in each participant, it is often considered advantageous to aggregate connectivity matrices from multiple participants into a composite matrix, thereby generating a more accurate estimate of temporally stable functional connectivity \citep{varoquaux2010brain, zuo2014test}. A disadvantage of the approach is that the derivation of a single composite matrix precludes an assessment of outcome variability. Here, we propose to assess outcome variability with respect to different instantiations of the composite matrix. We generated multiple estimates of the composite matrices for different age groups using a sub-sampling procedure, which allowed us to quantify the variability in our results and ultimately determine the robustness of our conclusions. This procedure is only possible due to the large number of participants. As neuroscience moves into the ``big data'' era this type of robustness testing will likely become more feasible and desirable \citep{zuo2014test}.

Second, we deal with community structure in a non-standard way. In many applications, a functional network's communities are considered to be the partition that optimizes some quality function (e.g. the $Q$ measure). However, it has been shown that the number of near-optimal solutions grows exponentially with the size of a network \citep{good2010performance}, making it unlikely that any modularity-maximization heuristic will uncover the globally optimal partition. It is unclear, then, why any single near-optimal solution should be preferred over any other near-optimal solution. The strategy we adopted here was to describe the statistics of an ensemble of near-optimal solutions. This approach is, perhaps, less satisfying in that it fails to resolve a single ``best'' community structure, but it allows assessing the robustness of communities across a distribution of near-optimal partitions, an approach that is less prone to error than one that depends upon a single instance of community structure.

\subsection*{Methodological Considerations}
As with any MRI study, there are a number of methodological considerations that one should take into account in interpreting these results. The first issue is related to subject head motion, which has been shown to produce artifactual correlation patterns in human fMRI analyses \citep{power2012spurious}, and is especially problematic when motion amplitude is correlated with a dependent variable, such as participant age \citep{satterthwaite2012impact}. We attempted to mitigate this concern by including pre-processing steps for reducing motion artifacts \citep{xu2015connectome} as well as regressing out motion parameters from variables of interest, such as modularity scores, and analyzing residuals (Figure \ref{fig:figs3}). While these steps help address issues related to head motion, it is also probable that they do not completely eliminate motion as a potential confound. Future development in pre-processing strategies for MRI data will likely help address this issue.

Another concern is related to our choice of node definition. It is well known that one's choice of nodes can have an influence on the properties of the resulting networks \citep{fornito2010network}. In the main text, we presented results wherein nodes were defined according to a so-called ``functional atlas'' \citep{yeo2011organization}. We also replicated our main findings by using a second parcellation, where nodes were defined according to anatomical landmarks \citep{destrieux2010automatic} (Figure \ref{fig:figs2}J-L).

A final concern is that the hemodynamic response (i.e. changes in blood volume, flow, and oxygen level) to neural activity varies across age groups \citep{d1999effect, d2003alterations}. In principle, such unwanted variation makes it difficult to ascribe changes in functional connectivity and community structure solely to changes in coordination between brain regions. Future work will undoubtedly help address this issue, as better, subject- and region-specific models of neurovascular coupling become available \citep{handwerker2012continuing}.

\section*{Conclusion}
The findings of our study support the conclusion that the community structure of the cerebral cortex undergoes age-related changes that unfold in characteristic patterns on multiple scales. The age-dependent evolution of functional communities in the brain is incompletely captured by describing changes on a single, coarse or fine, scale. The methods and approaches underpinning our analyses are likely to provide important additional information in uncovering variations in structural and functional networks across healthy and clinical populations.

\section*{Appendix}
In addition to the results presented in the main text, we performed a number of supplemental analyses to demonstrate the robustness of our results to variation in the number of age groups, cortical parcellation, and motion artifacts. This supplement details those analyses and also describes the process used to obtain the consensus communities described in the main text as well as the details of an additional null model against which we compared raw flexibility scores.

\subsection*{Robustness to variation in age groups}
\begin{figure}[ht]
\centering
\includegraphics[width=\linewidth]{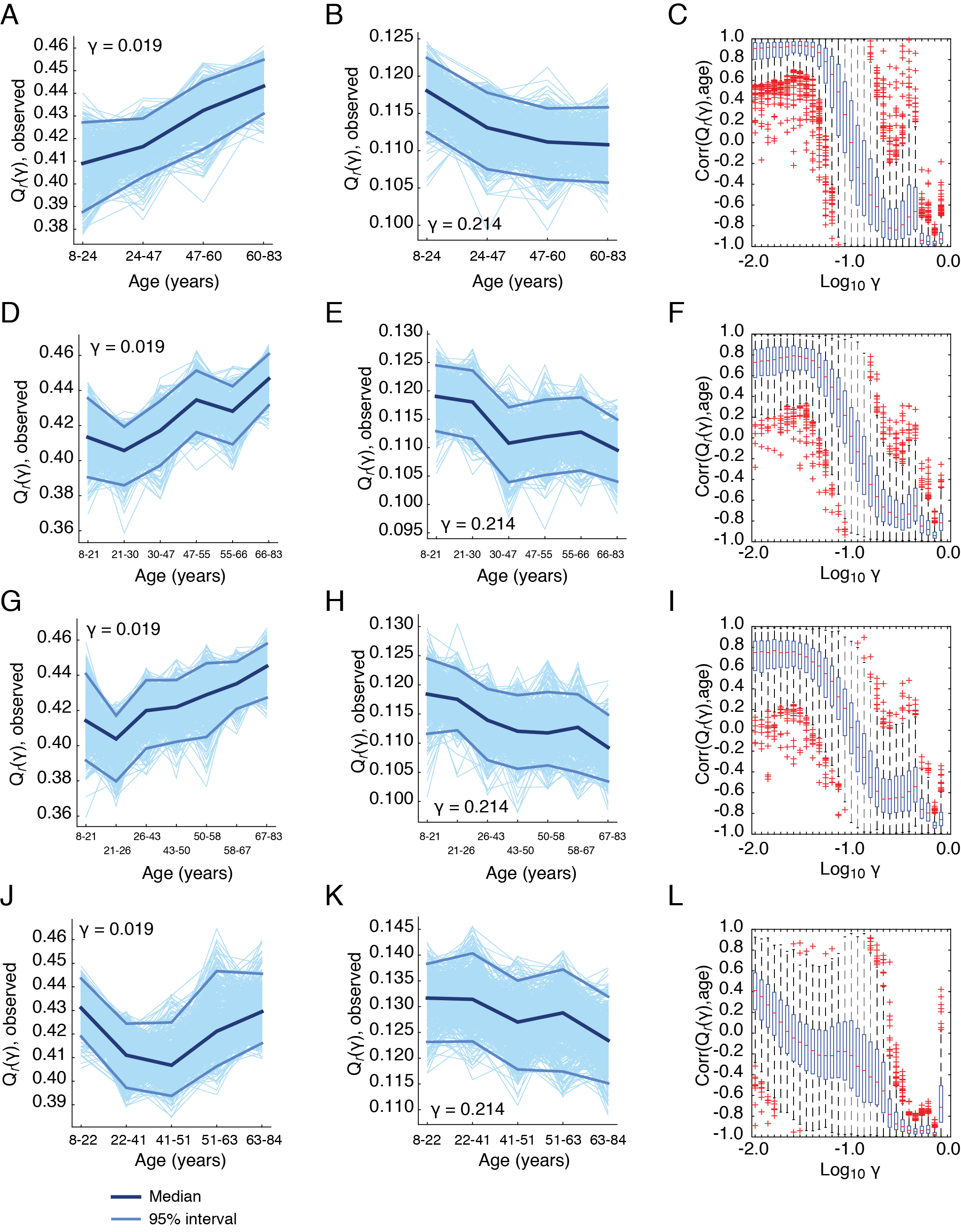}
\caption{Reproducing main results with different numbers of age groups and with different cortical parcellation. A-C) Scale specific changes in $Q_r(\gamma)$ when $K=4$. D-F) Scale specific changes in $Q_r(\gamma)$ when $K=6$. G-I) Scale specific changes in $Q_r(\gamma)$ when $K=7$. J-L) Scale specific changes in $Q_r(\gamma)$ when the cerebral cortex was parcellated according to the Destrieux atlas \citep{destrieux2010automatic}. Note that when using the Destrieux atlas, the previously observed linear relationship observed between age and $Q_r(\gamma)$ at coarse scales has become quadratic. As a consequence, linear correlation no longer does a good job describing this behavior. This explains the relatively weak correlation coefficients for $\log_{10}(\gamma) < -1.0$ in Panel L.}
\label{fig:figs2}
\end{figure}

\begin{figure}[ht]
\centering
\includegraphics[width=\linewidth]{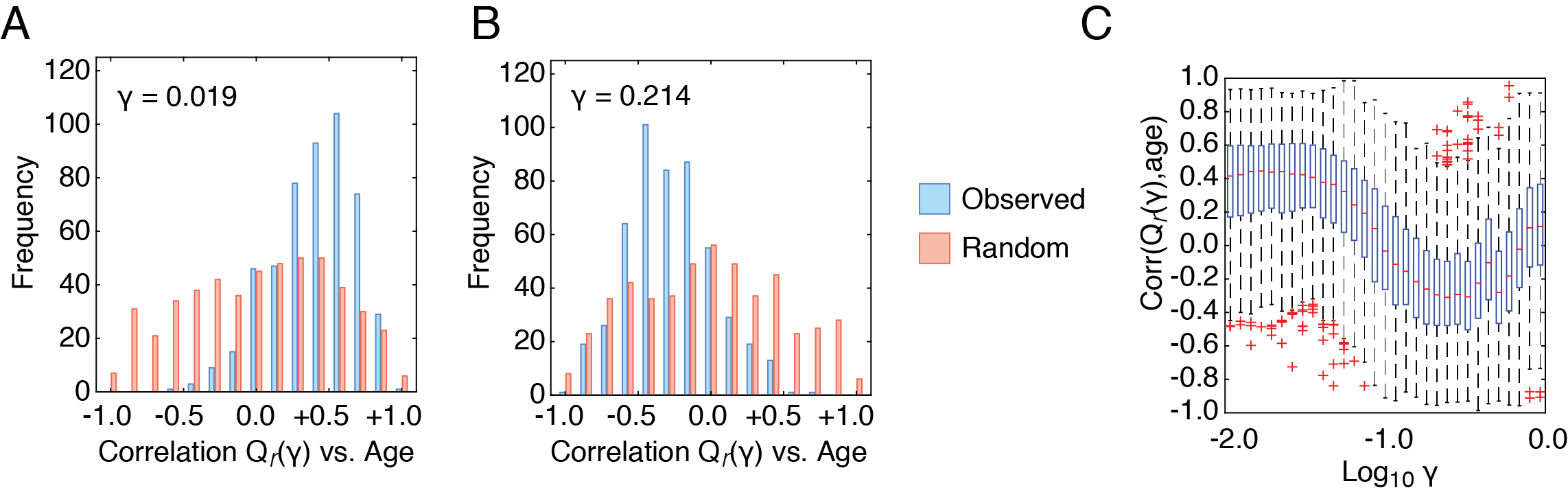}
\caption{Age versus modularity curves after performing additional motion correction step. A-B) Distributions of correlation coefficients, $\hat{r}_{age,Q_r(\gamma)}$, obtained from the observed data and \textit{network null model}, demonstrating the persistence of scale-specific modularity trajectories even after regressing out motion scores from the modularity scores. C) Distribution of correlation coefficients across all $\gamma$ values with fixed $\omega =0.032$.}
\label{fig:figs3}
\end{figure}

\begin{figure}[ht]
\centering
\includegraphics{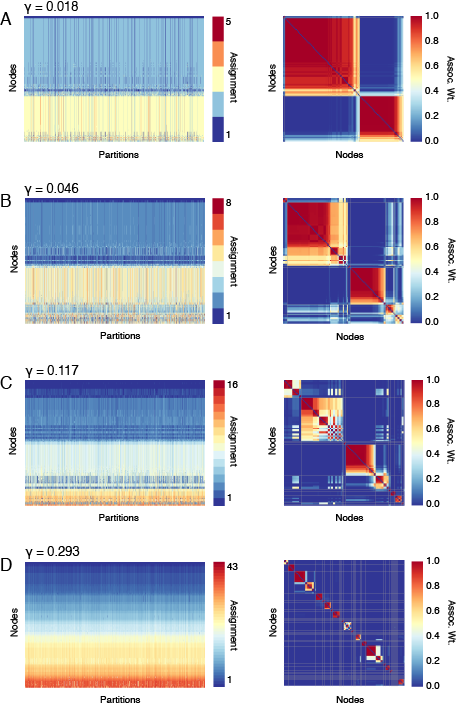}
\caption{Examples of detected single-layer partitions and consensus communities. A) Partitions detected with $\gamma = 0.018$ (five layers $\times$ 500 repetitions of the Louvain algorithm = 2500 total partitions) and association matrix ordered by consensus communities. B-D) Same as Panel A, but with $\gamma = 0.046$, $\gamma = 0.117$, and $\gamma = 0.293$. For all panels $\omega = 0.032$.}
\label{fig:figs4}
\end{figure}

In the main text we tracked the formation, evolution, and dissolution of modules across the human lifespan, focusing on $K=5$ age groups. Our principal finding was that at coarse scales (i.e. few communities) the segregation of communities, which we indexed as the single-layer modularity score, $Q_r(\gamma)$, increased with age, while at finer scales (i.e. greater number of communities) community segregation decreased. We sought to reproduce this result using different numbers of age groups, specifically when $K = \{4,6,7 \}$. In Figure \ref{fig:figs2}, we reproduce panels A and B from Figure \ref{fig:fig3} and panel A from Figure \ref{fig:fig4}. The first three rows of Figure \ref{fig:figs2} show age-related variation in $Q_r(\gamma)$ for numbers of age groups $K=\{ 4,6,7 \}$. In each row, the first two panels show $Q_r(\gamma)$ as a function of age for $\gamma = 0.019$ (i.e. Figures \ref{fig:figs2}A,D,G) and $\gamma = 0.214$ (i.e. Figures \ref{fig:figs2}B,E,H). The final panel in each row shows the distribution of correlation coefficients, $\hat{r}_{age,Q_r(\gamma)}$ for all values of $\gamma$ (Figures \ref{fig:figs2}C,F,I). Importantly, in all three panels we observe scale-specific variation in $Q_r(\gamma)$ similar to what was observed when $K=5$. Specifically, when $K=4$, the distribution of $\hat{r}_{age,Q_r(\gamma)}$ was significantly more positive than that of a null model at coarse scales (t-test, $df=998$, $t=32.46$) and significantly more negative than a null model at fine scales (t-test, $df=998$, $t=-20.46$). The same was true when $K=6$ (t-test, $df=998$, $t=32.92$ and $df=998$, $t=-26.73$) and $K=7$ (t-test, $df=998$, $t=34.60$ and $df=998$, $t=-29.52$).

\subsection*{Robustness to variation in cortical parcellation}
It is well known that the choice of cortical parcellation, which defines the nodes in a functional network, can bias graph-theoretic measurements made on the network. To help mitigate concern that the scale-specificity of $Q_r(\gamma)$ with age was not simply a product of our choice of parcellation, we re-analyzed our data using nodes defined by a different atlas \citep{destrieux2010automatic} and with the same $K=5$ age groups reported in the main text. Dividing the cortex according to this atlas resulted in $n=148$ nodes (74 per hemisphere). In Figure \ref{fig:figs2}J,K we show typical trajectories of $Q_r(\gamma)$ across the lifespan when $\gamma = 0.019$ and $\gamma = 0.214$, respectively. Note that at the coarse scale (Figure \ref{fig:figs2}J), the trajectory is no longer linear, and now follows an approximately quadratic trajectory. At the finer scale (Figure \ref{fig:figs2}K) the trajectory is qualitatively the same as that reported in the main text and in the previous section of this appendix (i.e. approximately linearly decreasing). The distribution of correlation coefficients for the coarse and fine partitions were significantly more positive and more negative what would be expected by change (t-tests, $df=998$, $t=7.83$, $p \approx 10^{-14}$ and $df=998$, $t=-18.47$, $p \approx 0$).

Because the coarse-scale trajectory is quadratic it is not useful to characterize it with a linear correlation coefficient as we did in Figure \ref{fig:fig4}A and Figures \ref{fig:figs2}C,F,I. Consequently, the distribution of correlation coefficients, $\hat{r}_{age,Q_r(\gamma)}$, does not favor large positive values at coarse scales (Figure \ref{fig:figs2}J). Nonetheless, the distribution exhibits clear scale-specific effects, consistent with the results presented in the main text.

\subsection*{Robustness to subject head motion}
It has become widely appreciated that head motion can introduce artifactual patterns of functional connectivity in fMRI studies \citep{power2012spurious}. To reduce such biases, we employed a state-of-the-art processing pipeline that included a cluster of measures for correcting head motion at both the individual and group levels (see description of data acquisition in earlier section) \citep{xu2015connectome}. An important question is whether the principal finding reported in the main text (the scale specificity of modularity with age) can be attributed to motion artifacts. Indeed, across the $N=316$ participants that we analyzed, motion (estimated as maximum frame-wise displacement) was modestly correlated with age ($\hat{r}_{age,motion} = 0.13$). To alleviate this concern, we estimated the average frame-wise displacement across the individual participants used to generate the group-average connectivity matrices (i.e. the matrices that formed the layers in our multi-layer representation). We then used linear regression analyses to orthogonalize the single-layer modularity scores $Q_r(\gamma)$ with respect to the motion estimates. The residual scores obtained following this regression were then correlated with age as before. We also repeated the same analysis for the matrices generated by random sampling of participants (i.e. the \textit{network null model}). As a result of this additional motion-correction step we found that the size of the effect attenuated, though we still found that $Q_r(\gamma)$ was positively correlated with age at coarse scales (median correlation of $\hat{r}_{age,Q_r(\gamma)}=0.41$ with interquartile range of $[ 0.18,0.57 ]$) and negatively correlated at fine scales (median correlation of $\hat{r}_{age,Q_r(\gamma)}=-0.31$ with interquartile range of $[ -0.47,-0.09 ]$). The observed distributions of correlation coefficients were significantly more positive (coarse scale, $df = 998$, $t=10.13$) and more negative (fine scale, $df=998$, $t=-8.20$) than one would expect under the random model.

\subsection*{Consensus communities}
As the size of a network grows, modularity maximization can yield exponentially many near-optimal partitions, making it difficult to identify the globally optimal partition \citep{good2010performance}. As a consequence, we tried to focus on the statistical property of the ensemble of near-optimal partitions rather than treating any single partition as representative. However, at times it was seen as advantageous to generate a single partition that, in some way, was representative of the partition ensemble. This process, more generally, is known as consensus clustering \cite{strehl2003cluster} and when applied to partitions of a network usually involves iteratively clustering an association matrix \citep{lancichinetti2012consensus}. An association matrix, $\mathbf{T}$, is a square $n\times n$ matrix whose element $T_{ij}$ represents the number or fraction of partitions in which nodes $i$ and $j$ were assigned to the same community across the entire partition ensemble. To obtain consensus communities from this partition, we re-cluster $T$ by finding the partition that maximizes the modularity $Q_{cons} = \sum_{ij} [T_{ij} - P_{ij}]\delta(g_i, g_j)$. Here, $P_{ij}$ is the probability of finding nodes $i$ and $j$ in the same community simply by chance. We obtained estimates of $P_{ij}$ by randomly permuting community assignments for each partition in the partition ensemble while preserving the number of size of communities. We maximized $Q_{cons}$ 500 times resulting in 500 estimates of consensus communities. Typically, the 500 consensus community estimates are identical (or nearly identical), in which case the clustering algorithm stops, having reached consensus \citep{bassett2013robust}. Otherwise, a new association matrix is generated from the consensus community estimates and the algorithm repeats until convergence. We show examples of partition ensembles and association matrices for $\gamma = \{0.018, 0.046, 0.117, 0.293\}$ and $\omega = 0.032$ in Figures \ref{fig:figs4}A-D. In all panels, nodes are ordered according to consensus communities.

\subsection*{Additional null model}
\begin{figure}[ht]
\centering
\includegraphics[width=\linewidth]{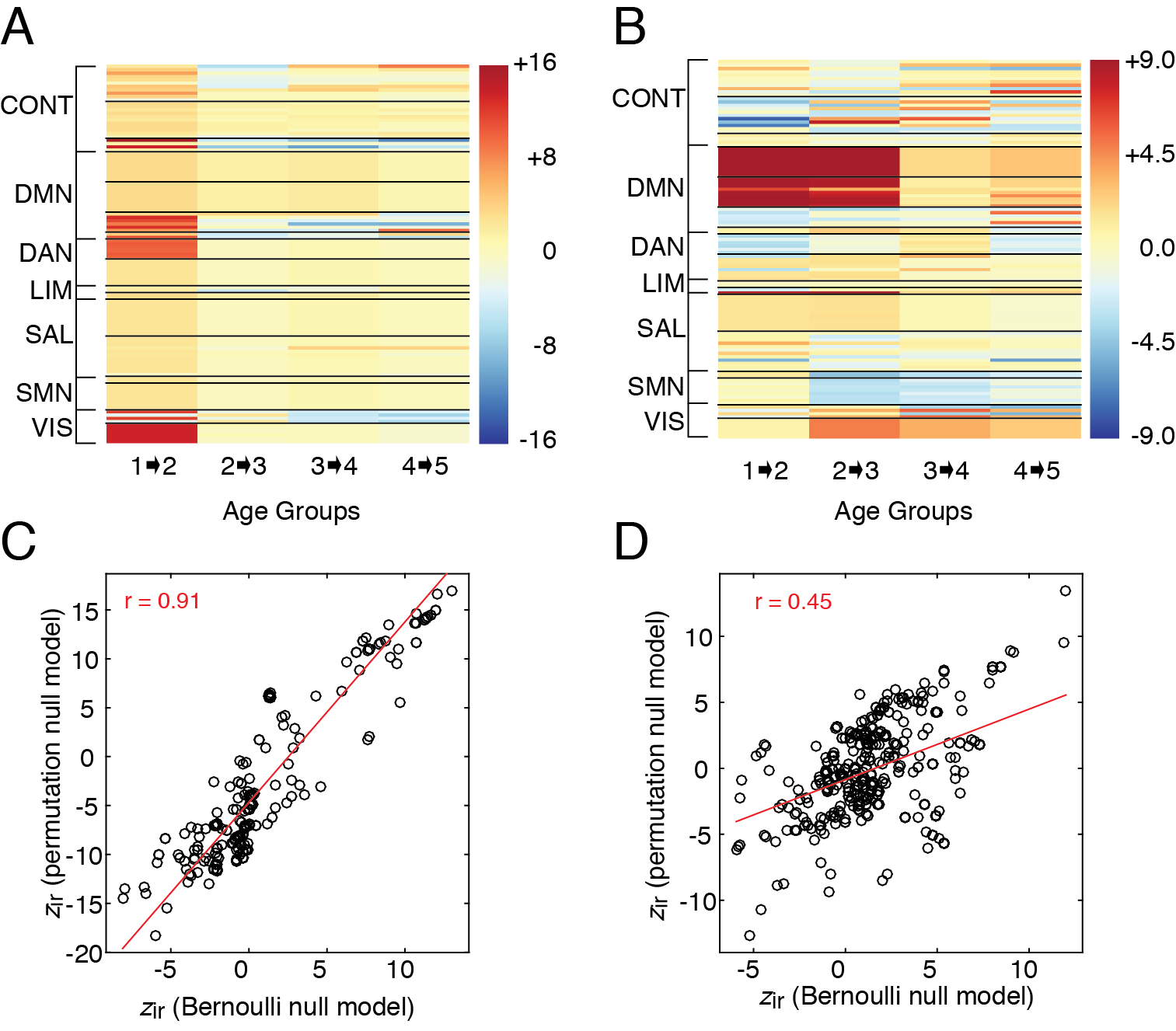}
\caption{Z-score regional flexibility scores estimated using an alternative null model. A) With resolution parameters $\gamma = 0.018$ and $\omega = 0.032$. B) With resolution parameters $\gamma = 0.214$ and $\omega = 0.032$. C-D) Scatter plots of z-score regional flexibilities estimated with the Bernoulli null model compared against those obtained from the permutation based null model for both $\gamma = 0.018$ and $\gamma = 0.214$.}
\label{fig:figs5}
\end{figure}

\begin{figure}[ht]
\centering
\includegraphics[width=\linewidth]{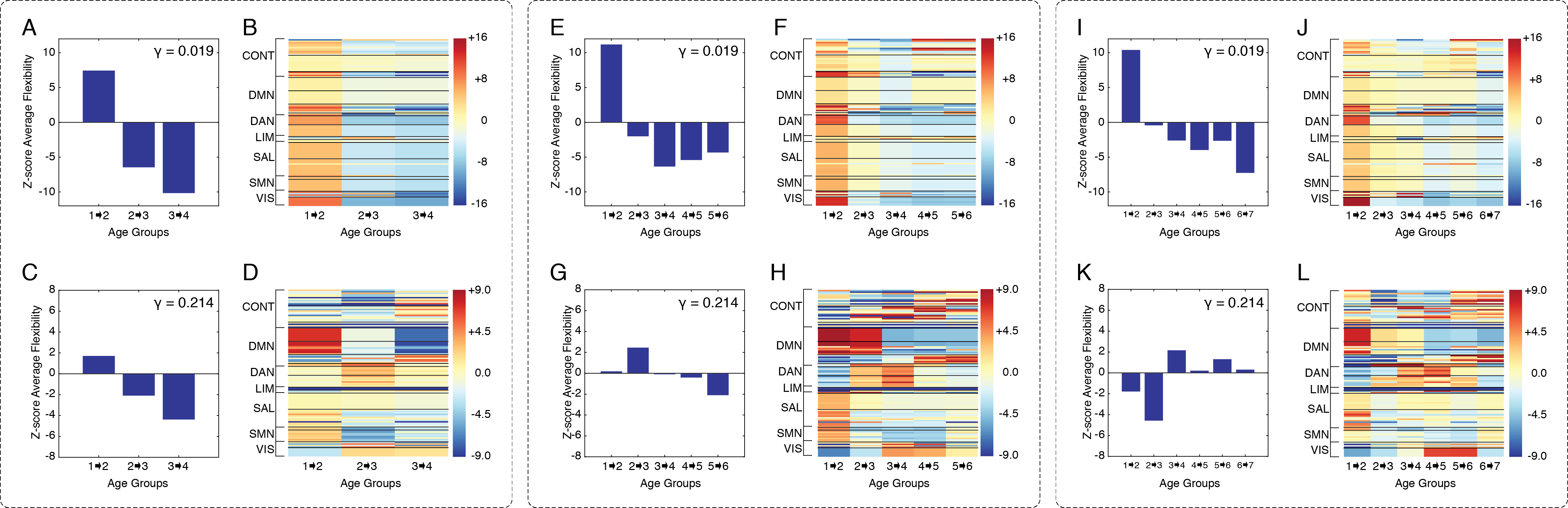}
\caption{Z-score regional flexibility scores estimated for $K=4$ (Panels A-D), $K=6$ (Panels E-H), and $K =7$ (Panels I-L). All plots were generated in precisely the same manner as those shown in Figure 6.}
\label{fig:figs6}
\end{figure}

In the main text we presented regional-level and average flexibility scores that we compared to the expected flexibility under a null model in which the community assignments of single-layer partitions was fixed but where the layer order was randomized. We also compared the raw flexibility scores to an additional null model. In this model we treat each node's flexibility score as a Bernoulli variable (i.e. there is some underlying probability that a node changes its community assignment between sequential layers). We compared the observed flexibility $f_{ir}$ to the randomized flexibility scores obtained from the \textit{network null model}, $f_{ir}^{rand}$ testing the null hypothesis that $f_{ir} = f_{ir}^{rand}$. The test statistic is given as:

\begin{equation}
z_{ir}^{rand} = \frac{f_{ir} - f_{ir}^{rand}}{\sqrt{p_{ir}(1 - p_{ir})\frac{2}{N_{reps}}}}
\end{equation}

where $p=\frac{f_{ir} + f_{ir}^{rand}}{2}$ and $N_{reps} = 500$ was the number of partitions in the partition ensemble. In general, we found results using this null model were similar to those obtained using the \textit{permutation null model}. We show the test statistics for $\gamma = 0.018$ and $\gamma = 0.214$ in Figures \ref{fig:figs5}A,B, respectively, and their correlation with the flexibility scores obtained from the permutation model shown in the main text (Figures \ref{fig:figs5}C,D).

\clearpage

\section*{References}
\bibliography{biblio}

\clearpage

\beginsupplement

\end{document}